# Formic Acid Decomposition Using Synthesized $Ag/TiO_2$ Nanocomposite in Ethanol-Water Media Under Illumination of Near UV Light

**E. Pipelzadeh[1,2], M. Valizadeh Derakhshan[1,2] , A. A. Babaluo[1,2,*], M. Haghighi[1,2], A. Tavakoli[1,2]**

1- Chemical engineering department, Sahand University of Technology, Tabriz, I.R. Iran
2- Nanostructure material research center (NMRC), Sahand University of Technology, Tabriz, I.R. Iran

(*) Corresponding author: a.babaluo@sut.ac.ir

## *Abstract:*
*The effect of ethanol-water media on the synthesis of $Ag/TiO_2$ nanocomposite was investigated with 0.05, 0.1 and 0.5 (wt.%) of Ag content. Ethanol was used as hole-scavenger enhancing the photodecomposition of $Ag^+$ ions under illumination of near-UV light. The nanocomposites were further calcined to 300 ˚C and 400 ˚C under controlled atmosphere. The synthesized nanocomposites were tested for photocatalytic efficiency decomposing formic acid as an organic pollutant under irradiation of a 4W near UV lamp with $l_{max}$ at 365nm and the results were compared with that of non-modified commercially available Degussa $TiO_2$ (P25). The synthesized nanocomposites were characterized using XRD and SEM micrographs. The photocatalytic studies have revealed a constant overall performance for all synthesized nanocomposites. The initial rate of decomposition was observed to increase with decreasing Ag content, in the case of 0.05 wt.% having the best initial rate. Calcination of the nanocomposites was found to have activating properties on the synthesized nanocomposites where nanocomposite with 0.05 (wt.%) Ag content calcined at 300°C was distinctively advantageous over other calcined nanocomposites.*

***Keywords:*** *Nanocomposite, $TiO_2$, Ag, Calcination, Photocatalytic Activity, Ethanol-water media*

## 1. INTRODUCTION

Many investigations has been conducted since 1972 when Fujishma et.al. first reported the water splitting capabilities of $TiO_2$ upon UV irradiation producing oxygen as a result [1]. This discovery has initiated investigations on various catalytic processes for environmental protection later known as photocatalysis. Photocatalytic materials has great potential to remove aqueous and air organic contaminates [2-4]. Titanium dioxide with Anatase, Rutile and Brookite crystal structures are believed to have several advantages over conventional oxidation processes, such as: (1) complete mineralization of the pollutants; (2) using the near-UV irradiation; (3) no addition of other chemicals; and (4) operation at near room temperature [5-7]. However, $TiO_2$ can be activated only under UV light with wavelengths less than 387nm irradiation due to its large band gap of 3.2 eV which leads to limit practical applications. The photocatalytic activity of $TiO_2$ usually depends on a competition between the following two processes, that is, the ratio of the transfer rate of surface charge carriers from the inner layer to the surface, and the recombination rate of photo-generated electrons and holes [8]. In order to improve its applications, different approaches have been investigated such as coupling with noble metals e.g. Ag [4, 9,10,21], Pt [4,9], Fe [8], Cu [11] or other semiconductors such as CdSe [12] and CdS [13], improving the photocatalyst sensitization in the visible region wavelengths and enhancing electron-hole separation [14]. Different methods such as ultrasonic wave [15], sol-gel [16,22] and photo reduction [9, 17,21] have been used for metal doping.

As the Fermi level of $TiO_2$ is higher than that of loaded silver, silver deposits behave as accumulation sites for photogenerated electrons transferred from $TiO_2$. When the number of silver clusters is small and the photogenerated electrons will transfer to silver clusters, better separation of electrons and holes will be achieved with the increase of silver loaded up to the optimum content [18,23]. These electrons could react with adsorbed molecular oxygen or surface $Ti^{4+}$ to form reactive species of $O_2$•− and reactive center surface $Ti^{3+}$ (OH-), respectively. Meanwhile the amount of recombination center of inner $Ti^{3+}$ decreases [10]. In addition, the loaded silver particles can transfer one electron to adsorbed molecular oxygen to form $O_2$•− or to the $TiO_2$ surface $Ti^{4+}$ to form surface $Ti^{3+}$ [10]. The presence of silver clusters would decrease the recombination rate while the generated $O_2$•− and $Ti^{3+}$ are found to increase that. This suggests that the recombination rate would decrease and the generation of $O_2$•− and surface $Ti^{3+}$ would be accelerated and the yield of $h^+$ would also be increased [18].

With increase of silver loading, the number and size of silver clusters become larger gradually and the energetic properties of the loaded silver may approach to that of bulk silver which makes the silver sites become the recombination center of photogenerated electrons and holes [18].

Doping of Ag at low concentrations and effect of further calcination under controlled atmosphere over doped $TiO_2$ and its effect on total mineralization of formic acid under UV illumination using a 4W-UV lamp with its $l_{max}$ at 365 nm is investigated in the present work. Silver is photo-reduced onto $TiO_2$ surface in an ethanol-water media photocatalytically, where the media is acting as hole-scavenger, enhancing the reduction of Ag ions to Ag metal. XRD and SEM micrographs have used for characterization of synthesized nanocomposites and their photocatalytic properties were compared with that of non-modified Degussa $TiO_2$ (P25) as a reference for measuring its enhanced properties.

## 2. EXPERIMENTAL

### 2.1. Materials

Titanium dioxide (P25) was purchased and used from Degussa, Germany, Formic acid, Ethanol 99.9% and Silver Nitrate (0.1M) was also purchased from Merck, Germany. For photocatalytic tests, deionized water was prepared.

### 2.2. Ag/$TiO_2$ Nanocomposite synthesis

$TiO_2$ (P25) (70% Anatase and 30% Rutile) was doped with silver atoms in an organic aqueous media, containing 0.05, 0.1 and 0.5 (wt. %) Ag with respect to $TiO_2$. Appropriate amount of $TiO_2$ was dispersed in 200ml of ethanol-water media and stirred for an hour under dark conditions. Silver nitrate was added as a silver ion source to the prepared $TiO_2$ suspension and mixed again for an hour under dark conditions. The prepared suspension was then subjected to a uni-wavelength 360nm UV lamp for 2 hours which was experimentally found suitable doping all added $Ag^+$ into metal Ag in a designed reactor as shown in Figure 1. Ethanol acts as a hole-scavenger providing the electrons required for the reduction

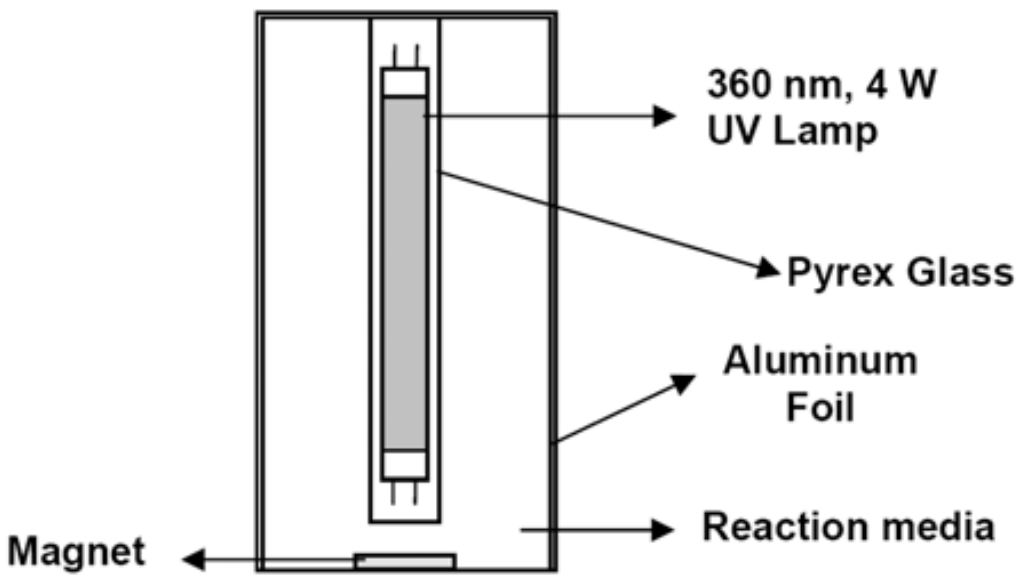

***Figure 1:*** *Reaction vessel for photocatalytic reduction of Ag on to $TiO_2$*

$$TiO_2 + h\nu \longrightarrow TiO_2\,(e^- + h^+) \quad (1)$$

$$O_2 + e^-\,(Ti^{III(3+)}) \longrightarrow O_2^{-\bullet} \quad (2)$$

$$OH^-_{surf}\ (\text{or } H_2O_{surf}) + h^+ \longrightarrow OH^{\bullet}_{surf} + (H^+) \quad (3)$$

$$C_2H_5OH + h^+ \longrightarrow C_2H_5O^-_{surf}{}^{\bullet} + (H^+)\ \text{(Dissociative chemisorption)} \quad (4)$$

$$C_2H_5O^-_{surf} + h^+ \longrightarrow C_2H_5O^{\bullet} \quad (5)$$

$$C_2H_5O^{\bullet} + h^+ \longrightarrow C_2H_4O_{surf} + H^+_{surf.} \quad (6)$$

$$C_2H_5O^-_{surf} + OH^{\bullet}_{surf} \longrightarrow C_2H_5O^{\bullet}_{surf} + OH_{-sur}f \quad (7)$$

$$C_2H_5O^{\bullet}_{surf} + OH^{\bullet}_{surf} \longrightarrow C_2H_4O_{surf} + H_2O_{surf} \quad (8)$$

$$C_2H_4O_{surf} + h^+/OH^{\bullet} \longrightarrow HCOOH \qquad CO + H_2O \quad (9)$$

$$e- + Ag^+ \longrightarrow Ag^{\circ} \quad (10)$$

$$Ag^{\circ} + Ag^+ + e^- \longrightarrow Ag^n \quad (11)$$

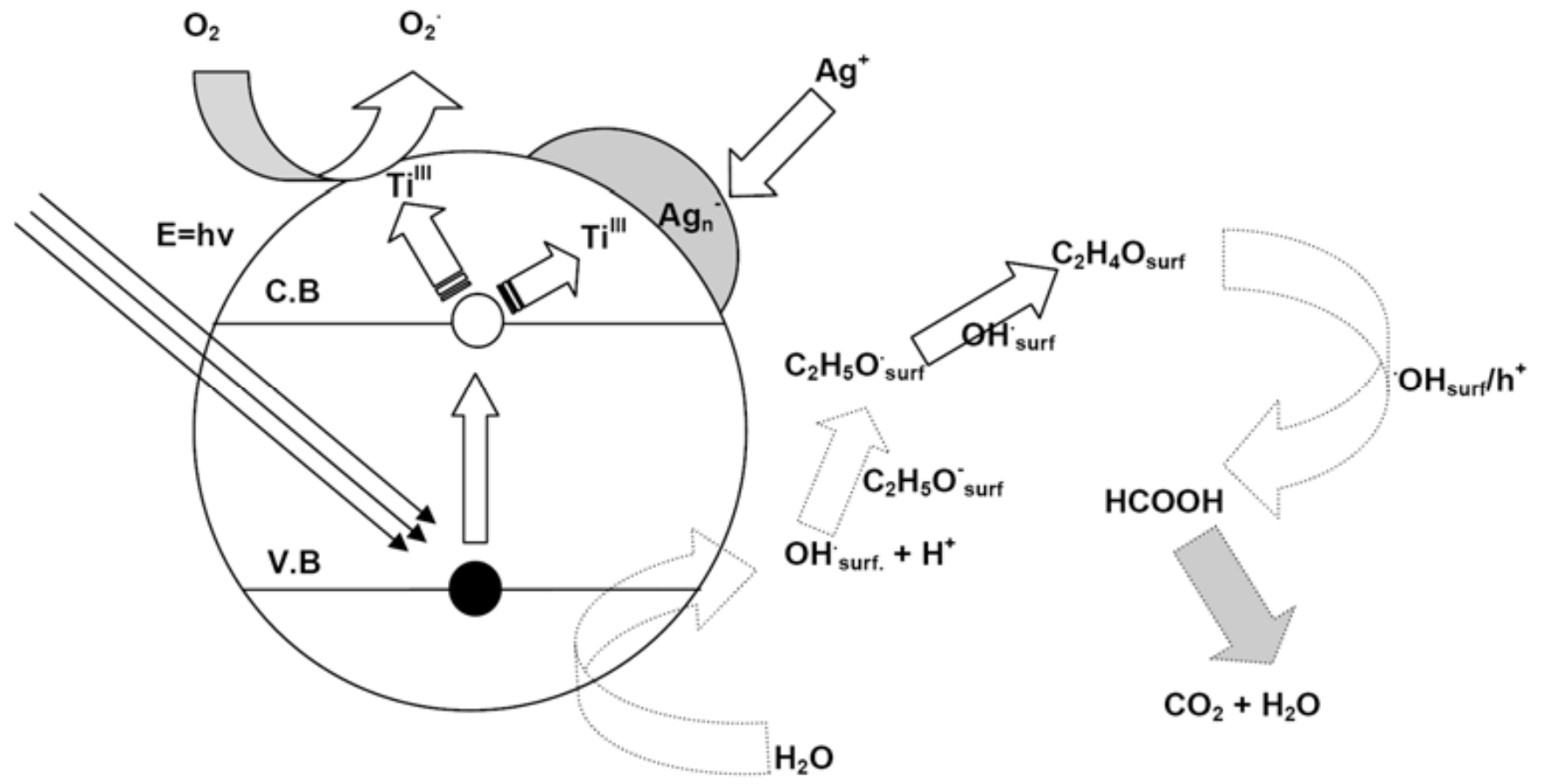

***Figure 2:*** *Proposed mechanism for the photoreduction of Ag onto* $TiO_2$ *in ethanol-water media*

of $Ag^+$ into Ag clusters. The slurry was filtered using a filter paper and washed with distilled water using a Buckner flask and dried at 60°C using an oven. The dried nanocomposite was hand-milled and subjected to a later calcination at 300°C and 400°C using a furnace under argon-controlled atmosphere. Figure 2 shows the photoreduction mechanism of $Ag^+$ ions onto $TiO_2$ substrate in an ethanol-water media where ethanol is acting as a hole-scavenger. Mechanism of ethanol decomposition can be proposed as equations (1) - (9) and demonstrated schematically in Figure 2 based on the literature [19, 21].

### 2.3. Characterization

XRD patterns were obtained using TW3710 Philips X'Pert diffractometer using CuKα as radiation and CuKα as a filter (λ=1.54 Å). Data were collected in the range of $20 \leq 2\theta^o \leq 90$ a $0.02^o$ 2θ-step and 2 sec per step (40 kV and 30 mA). Morphological studies were conducted using a scanning electron microscopy (SEM, Viga II, $3x10^5$, USA).

### 2.4. Photocatalytic performance tests

Formic acid was taken as the organic pollutant in the aqueous media subjected to either of the above synthesized nanocomposites and that of the untreated $TiO_2$ .5μl of 1M formic acid was added to 0.1 g of the nanocomposite suspension in 100 ml of deionized water using a 250 ml beaker and irradiated with a 4 W (uni-wavelength UV lamp with its $\lambda_{max}$ at 365nm) in a pre-designed reactor and mixed using a magnetic stirrer, and its decomposition was tracked using a digital pH meter. Figure 3 schematically shows the setup used for the photocatalytic performance test.

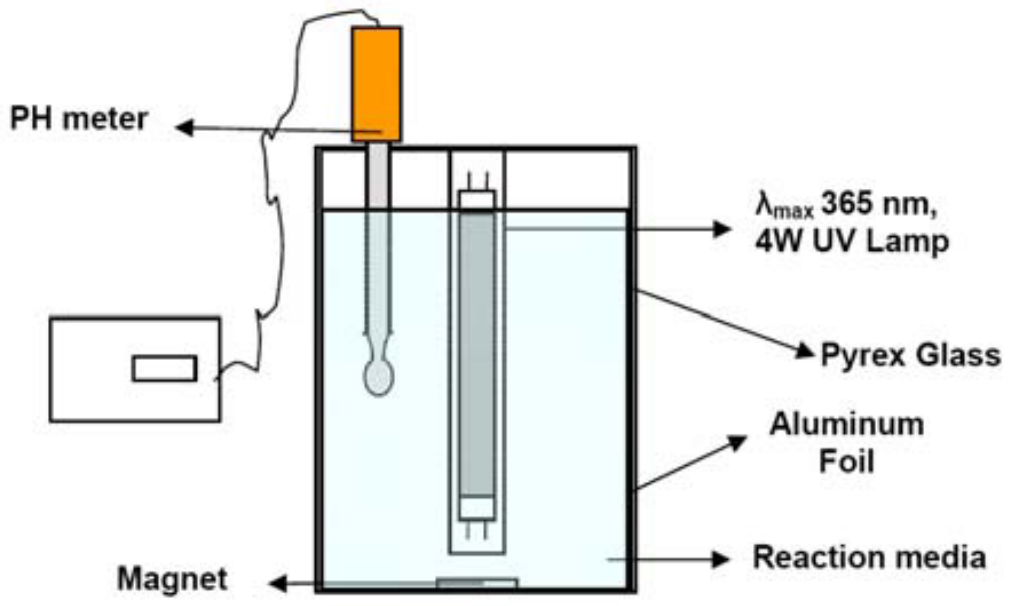

***Figure 3:*** *Photocatalytic performance setup using a 4W Uni-wavelength UV lamp ($\lambda_{max}$ 365nm)*

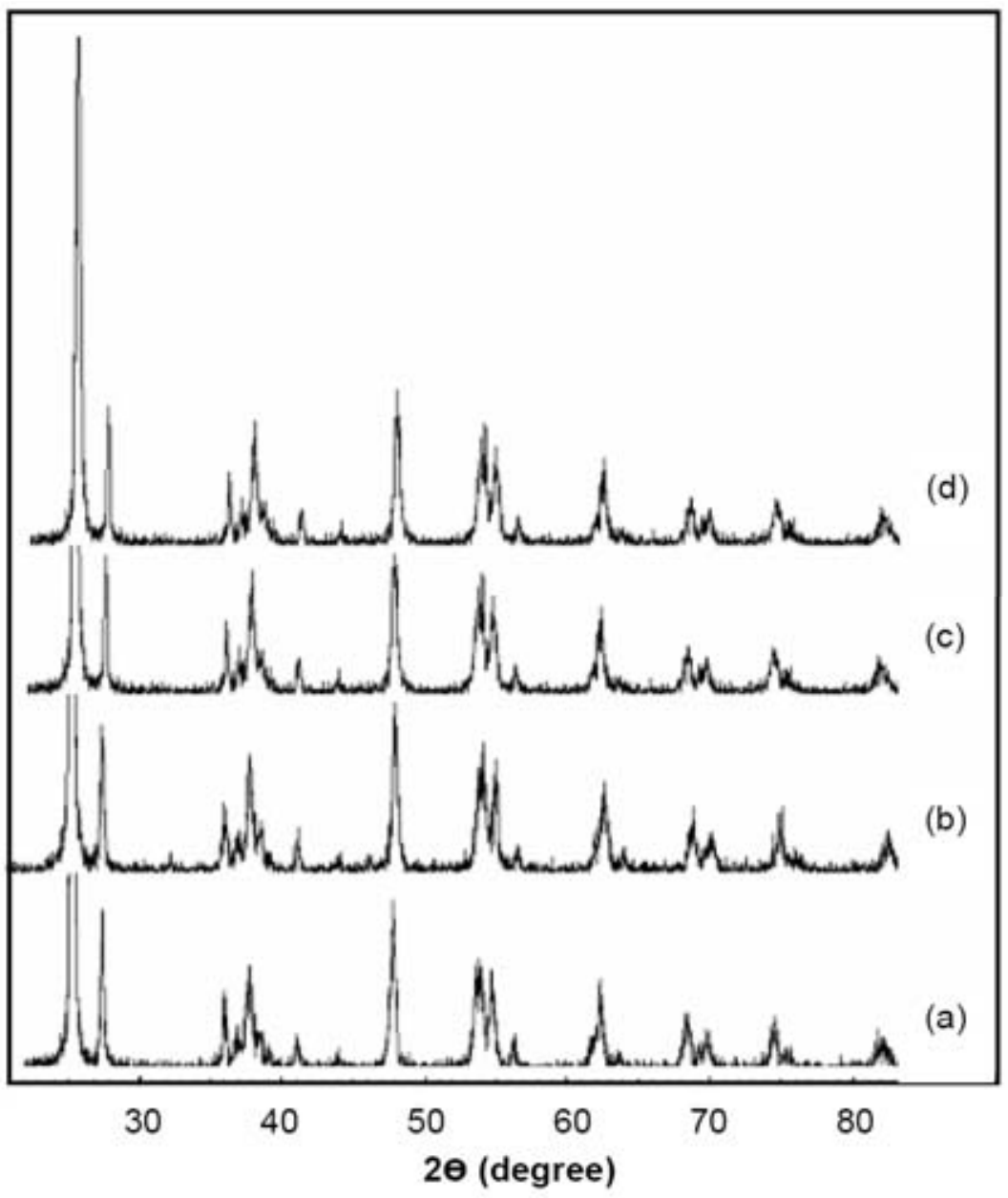

***Figure 4:*** *XRD patterns of the Ag/$TiO_2$ nanocomposites synthesized using different amounts of doped Ag: (a) 0.05, (b) 0.1 and (c) 0.5 (wt.%) Ag/$TiO_2$ nanocomposites and (d) $TiO_2$ (P25)*

## 3. RESULTS AND DISCUSSIONS

### 3.1. Crystallographic characterization

Figure 4 shows the XRD patterns of the Ag/$TiO_2$ nanocomposites synthesized using different amounts of doped Ag. No significant difference is observed between XRD results obtained from the synthesized nanocomposites compared with that of obtained from non-modified $TiO_2$ (Degussa, P25). The peaks in the XRD patterns correspond to the Anatase and Rutile phases, from the commercial $TiO_2$. No sign of crystal Ag is observed in the XRD analysis obtained from both the calcined and non-calcined samples at all Ag doped concentrations, suggesting either that the Ag is not in the crystal form or the extent of deposition is not detectable by XRD analysis.

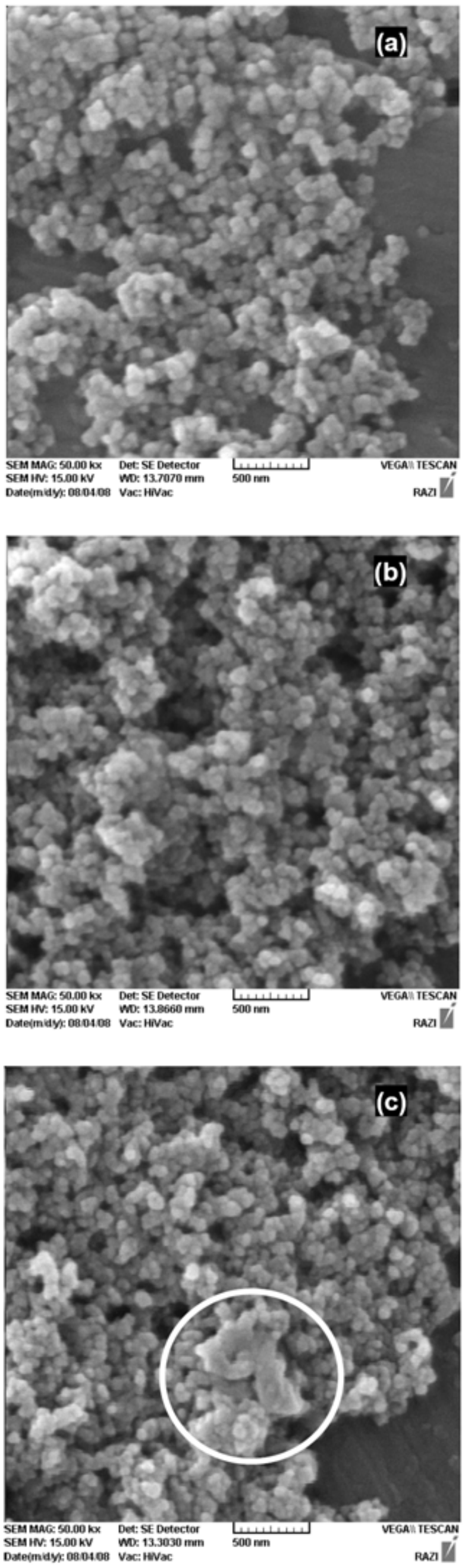

***Figure 5:** SEM imagines of (a) 0.5% $Ag/TiO_2$, (b) 0.5% $Ag/TiO_2$ 300 °C calcined and (c) 0.5% $Ag/TiO_2$ 400 °C calcinedFigure 6. Elemental analysis of 0.5 (wt.%) $Ag/TiO_2$ nanocomposite*

### 3.2. Surface Morphological Characterization

Figure 5 (a-c) shows the micrographs of 0.5 (wt. %) $Ag/TiO_2$ nanocomposites in both non-calcined and calcined states. As it is clearly indicated from the presented micrographs, the calcined form is more aggregated and growth of particles are seen as a result of sintering at 400°C, reducing active surface area and effecting the size distribution. Calcined $Ag/TiO_2$ nanocomposites at 300°C shows less aggregation than the 400°C and better sizing distribution which is also in good agreement with literature [20]. Figure 6 shows the results of elemental analysis for 0.5 (wt. %) $Ag/TiO_2$ nanocomposite calcined at 300°C, revealing both the presence of metallic silver on the surface of the $TiO_2$ nanoparticles and the chemical composition of the silver with respect to titanium atoms. The obtained composition is in good agreement with the experimental data.

### 3.3. Photocatalytic performance studies

Figure 7 (a) shows photocatalytic performance of the synthesized $Ag/TiO_2$ nanocomposites and compared with that of commercially available Degussa $TiO_2$ (P25) under illumination of near UV light. The rate constants "K" based on first order kinetics equation of (Ln C/C0= - Kt) photodegradation were calculated and presented in Table1.

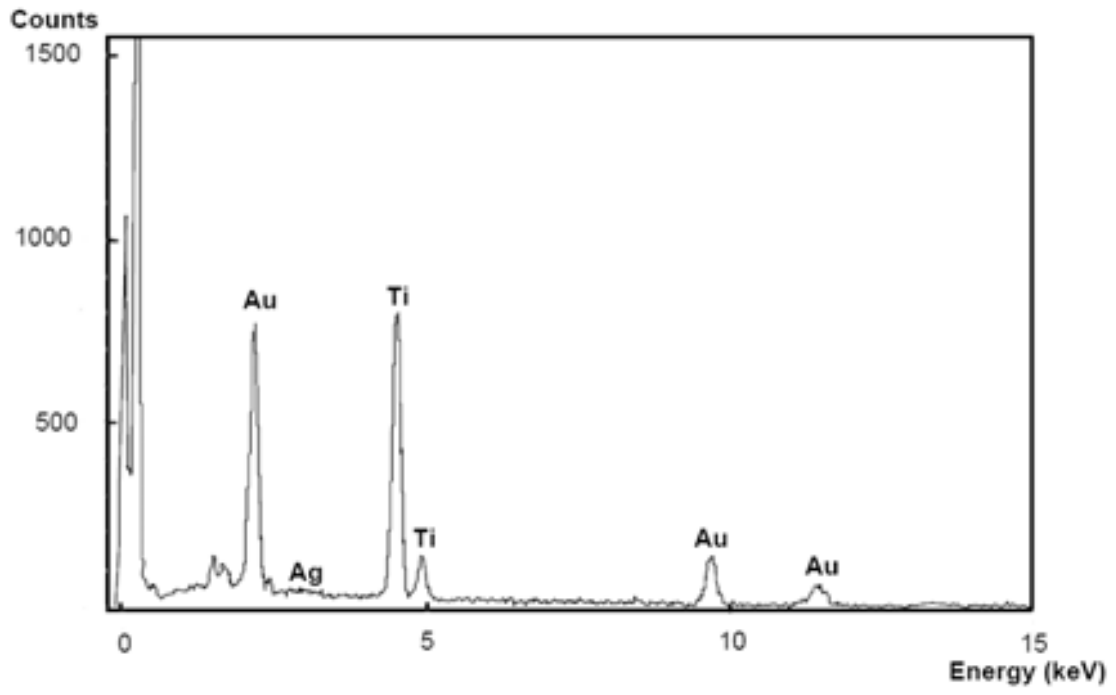

***Figure 6:** Elemental analysis of 0.5 (wt.%) $Ag/TiO_2$ nanocomposite*

Figure 8 (a) and (b) schematically shows the photodecomposition path of formic acid in the presence of non-modified $TiO_2$ and $Ag/TiO_2$ nanocomposite respectively. The results show an improved photocatalytic performance for the synthesized $Ag/TiO_2$ nanocomposites compared with that of non-modified $TiO_2$ (P25). The results reveal no significant difference in the time required for total organic compound decomposition for 0.05 and 0.1 wt.% $Ag/TiO_2$ nanocomposites but the difference in the initial decomposition rates is considerable where the former (0.05 (wt. %) $Ag/TiO_2$) shows the best initial rate in the investigated range. This initial decomposition rate is decreased with increase in Ag content. Results for 90% photodecomposition of the organic compound clearly shows the significant advantage of 0.05 (wt. %) $Ag/TiO_2$ over other nanocomposites and that of non-modified $TiO_2$ (P25) revealing that 0.05 wt.% Ag content has shown the best photocatalytic performance under near-UV illumination.

The doped Ag is found to act as an electron sink; storing photogenerated electrons providing a longer time for electrons to react with the electron accepting species such as oxygen molecules in the aqueous media.

Adsorbed ethanol was found to provide the necessary spacing between the doped Ag due to its molecular size which is decomposed to acetaldehydes and formaldehydes upon oxidation. The adsorbed ethanol and its decomposed forms such as formaldehydes which are produced in the course of ethanol decomposition are found to prevent the formation of uniform Ag layers or clusters that would in return restrict the light from entering the $TiO_2$ structure. Such adsorbed organic intermediates had to be decomposed before any photocatalytic tests could be performed; for this, the nanocomposites were suspended in an aqueous media and irradiated with near-UV light until its pH value came to a distinct equilibrium. The initial Ag seeds are found to be formed on the high potential sites having strong desire to donate electron, such sites are $Ti^{III}$-rich sites which have accepted an electron from the $TiO_2$ bulk and are near enough to the surface to reduce an $Ag^+$ ion into metal $Ag^o$.

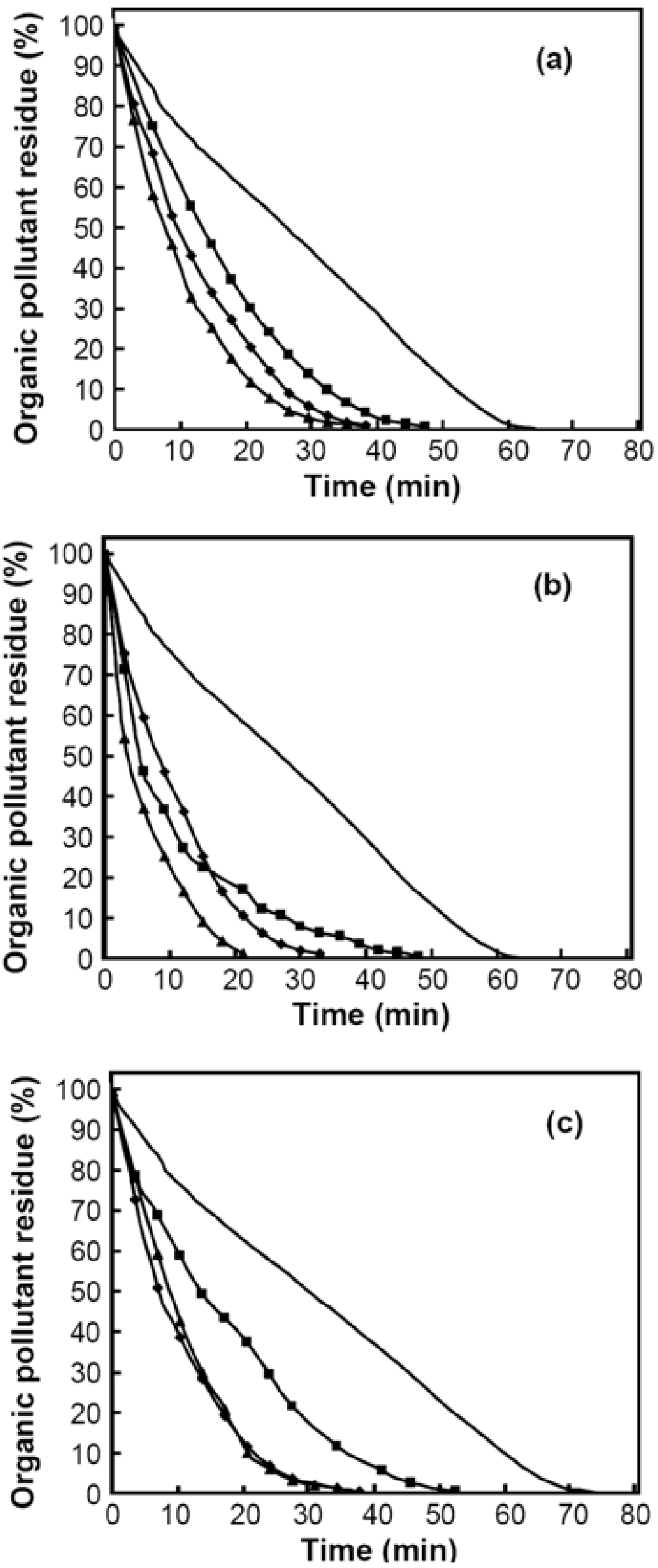

***Figure 7:*** *Photocatalytic activity of $TiO_2$ and $Ag/TiO_2$ nanocomposites: (▲)0.05 wt.% $Ag/TiO_2$ (■) 0.1 wt.% $Ag/TiO_2$ (◆) 0.5 wt.% $Ag/TiO_2$ (✖) Non-modified $TiO_2$ (P25), (a)non-calcined, (b)$Ag/TiO_2$ 300°C and (c) $Ag/TiO_2$ 400°C.*

***Table 1:*** *Rate constants "K" based on first order kinetics equation of (Ln C/C0= - Kt) photodegradation*

| No. | Ag (wt%) | Calcination | K ($min^{-1}$) |
|---|---|---|---|
| 1 | (▲)0.05 wt.% | noncalcined $TiO_2$ | 0.111 |
| 2 | (◆) 0.1 wt.% | | 0.082 |
| 3 | (■) 0.5 wt.% | | 0.064 |
| 4 | (x) Non-modified | | 0.037 |
| 5 | (▲)0.05 wt.% | 300°C | 0.174 |
| 6 | (◆) 0.1 wt.% | | 0.122 |
| 7 | (■) 0.5 wt.% | | 0.078 |
| 8 | (x) Non-modified | | 0.033 |
| 9 | (▲)0.05 wt.% | 400°C | 0.143 |
| 10 | (◆) 0.1 wt.% | | 0.139 |
| 11 | (■) 0.5 wt.% | | 0.079 |
| 12 | (x) Non-modified | | 0.049 |

Ag clusters are assumed to have more of a metallic character when exceeds some limits, not only has good electron-accepting capabilities from the $TiO_2$ bulk, but on the other hand, recombine the stored electrons with the holes decreasing their photodecomposition efficiency; this would explain the less photocatalytic activity observed from the synthesized 0.5 (wt.%) Ag/$TiO_2$ nanocomposite.

### 3.4. Effect of calcination on photocatalytic activity

The synthesized nanocomposites were further calcined up to 300°C and 400°C under controlled atmosphere. The calcined nanocomposites were suspended in aqueous media and illuminated with near-UV light for the test of photocatalytic activity. Figure 7 (b) and (c) show the results from the calcined nanocomposites. Calcination has shown to have activating effect on the 0.05 (wt. %) Ag/$TiO_2$ nanocomposite but less effect on other 300$^{oC}$ calcined nanocomposites. Calcination is assumed to cause migration of photoreduced metal silver forming larger Ag clusters on the $TiO_2$ surface. The organic contaminates adsorbed as a result of ethanol oxidation are evaporated by heating and such clusters are assumed to be readily formed. The decrement in photocatalytic activity of the calcined nanocomposites are found to be as a result of Ag clusters migration, forming larger clusters and blocking the UV light from entering the $TiO_2$ structure. Sintering of nanocomposites as a result of calcination is observed to form large aggregates of particles resulting in growth of particles reducing the specific surface area of the nanocomposites. The doped Ag is assumed to act as a soft media on the tough $TiO_2$ substrate causing such aggregates to form.

## 4. CONCLUSION

$TiO_2$/Ag nanocomposites were prepared using a chemical wet approach where ethanol-water media was used for the photoreduction of Ag onto

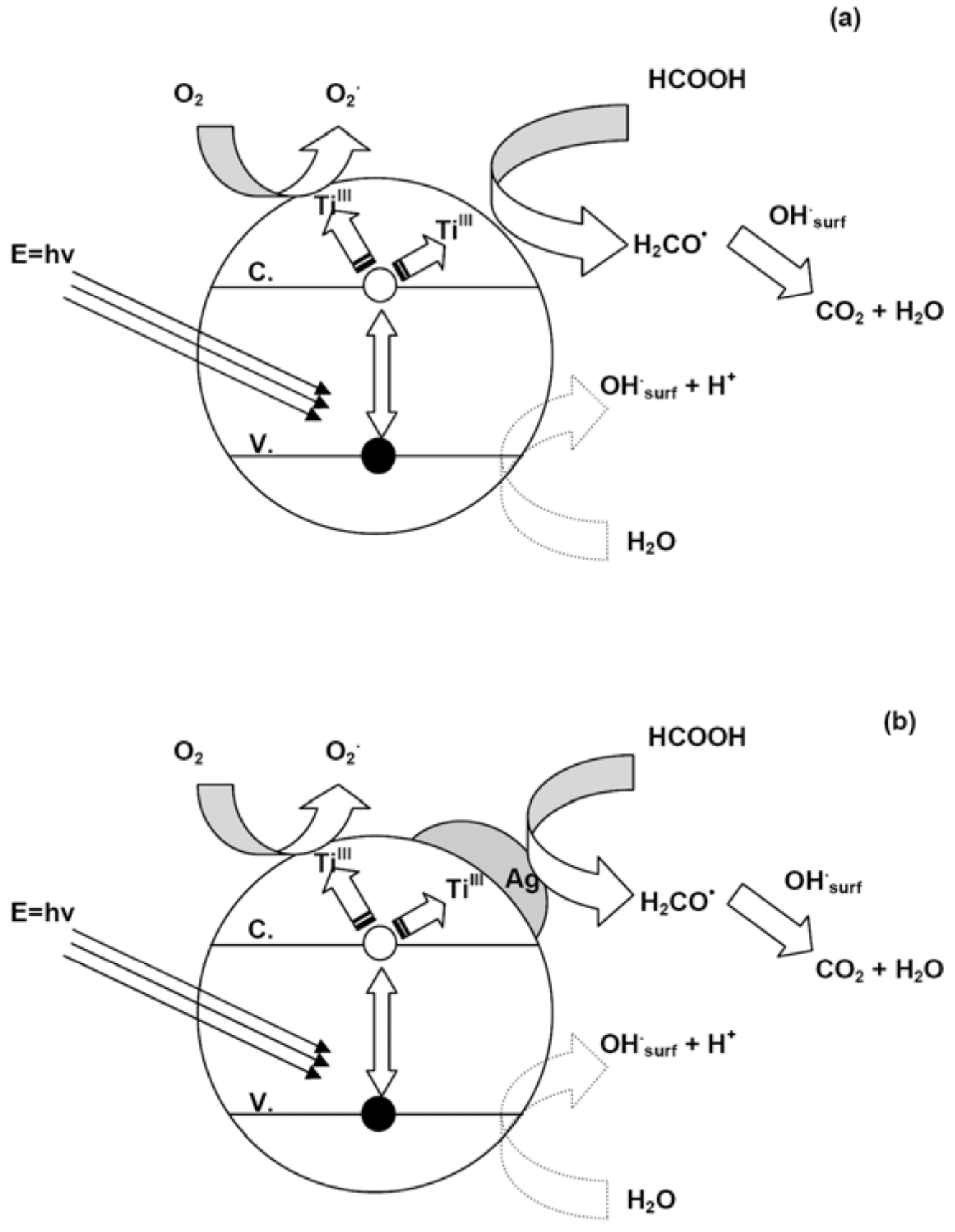

***Figure 8:*** *Schematic of the photocatalytic photodecomposition path of formic acid, (a) $TiO_2$ (b) Ag/$TiO_2$ nanocomposite*

$TiO_2$ under illumination of near-UV lamp with its $\lambda_{max}$ at 365nm. 0.05, 0.1 and 0.5 (wt.%) Ag-$TiO_2$ nanocomposites were prepared and further calcined up to 300°C and 400°C under controlled atmosphere. The fallowing conclusions can be made:

1- Ag doping was successfully accomplished in the ethanol-water media where ethanol was used as a hole-scavenger, increasing the rate of Ag photodeposition. The photocatalytic activity of the synthesized Ag/$TiO_2$ nanocomposites was found to be better than non-modified Degussa $TiO_2$ (P25).

2- Crystallographic analysis revealed that Ag doping does not affect the crystal structure of the $TiO_2$ as no sign of Ag is seen which may be due to either low Ag concentrations that could not be detected or that Ag did not have crystal forms; suggesting advances in the photocatalytic activity be totally contributed to the doped Ag.

3- 0.05 (wt. %) Ag/$TiO_2$ has shown the best initial photocatalytic activity which was seen to decrease as the Ag content increased.

4- The adsorbed ethanol and its oxidized organic byproducts are suggested to prevent the Ag seeds from growing to form uniform layers resulting in similar photocatalytic activity prior to calcination from the synthesized $Ag/TiO_2$ nanocomposites.

5- Calcination has shown to have activating impact on 0.05 (wt. %) $Ag/TiO_2$ nanocomposite but its influence was less considerable in comparison with other nanocomposites as result of aggregates formed reducing its specific surface area.

## ACKNOWLEDGEMENT

The authors wish to thank Prof. Pourabbas for providing the facilities at the NMRC and Sahand University of Technology for the financial support of this project.